# COMPARISON OF THE TRANSPORT MECHANISM IN UNDERDOPED HIGH TEMPERATURE SUPERCONDUCTORS AND IN SPIN LADDERS

JOHAN VANACKEN, LIEVEN TRAPPENIERS, GERD TENIERS, PATRICK WAGNER, KRIS ROSSEEL, JOEL PERRET*, JEAN-PIERRE LOCQUET**, VICTOR V. MOSHCHALKOV AND YVAN BRUYNSERAEDE

*Laboratorium voor Vaste-Stoffysica en Magnetisme, Katholieke Universiteit Leuven, Celestijnenlaan 200 D, B-3001 Leuven*
* *Institut de Physique, Université de Neuchâtel, CH-2000 Neuchâtel, Switzerland*
** *IBM Research Division, Zurich Research Laboratory, CH-8803 Ruschlikon, Switzerland*

Recently, the normal state resistivity of high temperature superconductors (in particular in $La_{2-x}Sr_xCuO_4$ single crystals) has been studied extensively in the region below $T_c$ by suppressing the superconducting state in high magnetic fields[1]. In the present work we report on the normal state resistance of underdoped $La_{2-x}Sr_xCuO_4$ thin films under epitaxial strain[2], measured far below $T_c$ by applying pulsed fields up to 60 T[3]. We will compare the transport measurements on these high temperature superconductors with transport data reported for the $Sr_{2.5}Ca_{11.5}Cu_{24}O_{41}$ spin ladder compound[4]. This comparison leads to an interpretation of the data in terms of the recently proposed 1D quantum transport model and the charge-stripe models[5,6]

## 1 Introduction

In order to understand the origin of superconductivity in the high temperature superconductors (HTSC) materials, one first needs to know the underlying normal state properties of these materials at temperatures T -> 0. The latter is hidden, however, behind the superconducting state. Nevertheless, the underlying normal state can be probed by using high magnetic fields. The most commonly used method to achieve these high fields, is the pulsed field technique[3]. This approach has been very successfully applied by Ando et al.[1] to investigate the ground state properties of $La_{2-x}Sr_xCuO_4$ single crystals with different Sr contents. Recently, it has been shown that the critical temperature $T_c$ of the La-Sr-Cu-O compound can also be tuned by changing the lattice mismatch between the substrate and the $La_{1.9}Sr_{0.1}CuO_4$ ultra thin film[2]. The comparison of the high field transport properties of strained and regular samples is one of the basic motivations for this work.

## 2 Experimental results and discussion

**Material preparation and characterisation:** The advantage in using strained $La_{2-x}Sr_xCuO_4$ thin films, is that for a fixed stoichiometry, the lattice dimensions can be changed (enlarged or decreased), thus strongly affecting the critical temperatures $T_c$. In this work, we present data on three samples with $x = 0.1$: Sample A and B are respectively 125 and 150 Å thin films, grown on $SrLaAlO_4$ whereas sample C is a 150 Å thin film grown on $SrTiO_3$. Using $SrLaAlO_4$ as a substrate leads to compressive strain, since the lattice parameters of $SrLaAlO_4$ are smaller than those of $La_{1.9}Sr_{0.1}CuO_4$. The inverse happens when growing films on $SrTiO_3$: tensile stress is induced due to the larger in-plane parameters of $SrTiO_3$. Figure 1(a) shows the different values of the ab-plane lattice parameters.

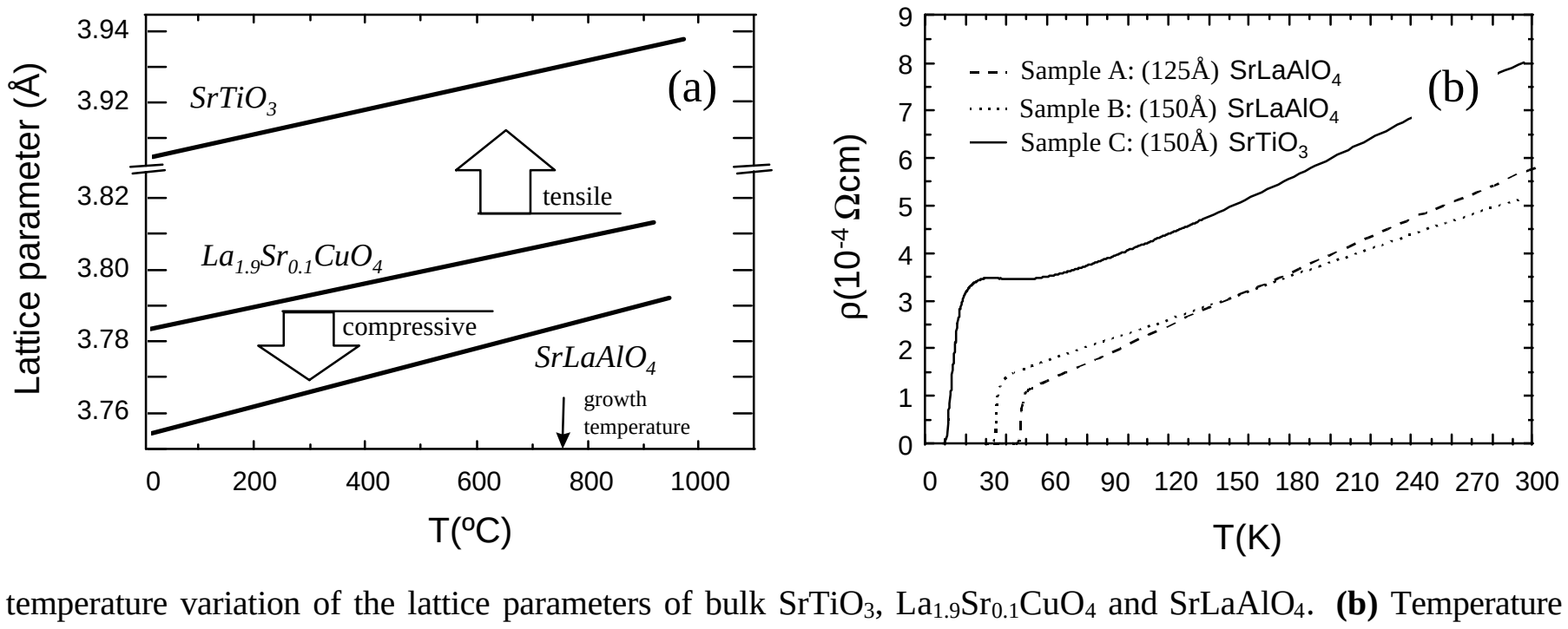

**Figure 1. (a)** The temperature variation of the lattice parameters of bulk $SrTiO_3$, $La_{1.9}Sr_{0.1}CuO_4$ and $SrLaAlO_4$. **(b)** Temperature dependence of the resistivity for different $La_{1.9}Sr_{0.1}CuO_4$ samples (A, B, C).

The influence of the stress induced by the substrate on $T_c$ can be seen from figure 1 (b). By applying tensile stress the ab-lattice parameters increases and the c-axis lattice parameter decreases, a process which leads to reduced $T_c$ values. By applying compressive stress, the opposite structural effects occur, leading to an increase of $T_c$.

**Normal state transport mechanisms:** The temperature dependence of the resistivity $\rho_{ab}$ in these thin films can be discussed in terms of three different dependencies. The *first* one, which is always present at high temperatures is a ***linear law*** ($\rho \propto T$) and has already been shown in the first measurements on optimally high temperature superconductors. It is even commonly used to prove the good quality of HTSC's. *Secondly*, we observed a ***super-linear*** behavior, which has been described by a square dependency in the past. We will argue, that this super-linear behavior is caused by the opening of the spin gap[4]. This second behavior, is barely present in the optimally doped samples, but becomes very pronounced in underdoped HTSC samples. The *third* dependency is only visible in the underdoped samples and at very low temperatures. In this case, the resistance ***diverges logarithmically*** with the decreasing temperature. We usually observe the opening of the spin gap, and the logarithmic divergence together in the same sample.

The model which we use to describe the measurements has been reported previously[5, 6], and its results are summarized in the table below. The model is based on the following basic principles. The Cu-O HTSC's, all have $CuO_2$ planes as building blocks. Such an undoped $CuO_2$ plane is a 2D antiferromagnet. When holes are introduced in the $CuO_2$ plane, the AF groundstate will be perturbed. In the case of low doping, it will be very difficult for the holes to become mobile, and they will be localized, surrounded by an AF background. With larger amounts of holes, there will probably be a phase segregation between charge areas, and AF areas. It is however clear, that all charge transport must be strongly influenced by magnetic scattering. Therefore, we[5] proposed that (i) the dominant scattering mechanism is of magnetic origin; (ii) the resistivity is determined by the inverse quantum conductivity $\sigma^{-1}$; and (iii) the inelastic scattering length $L_\phi$ is controlled by the magnetic correlation length $\xi_m$ via the strong interaction between the carriers and the $Cu^{2+}$ spins. The exact expression of the magnetic correlation length $\xi_m$ is dependent on the effective dimensionality of the system.

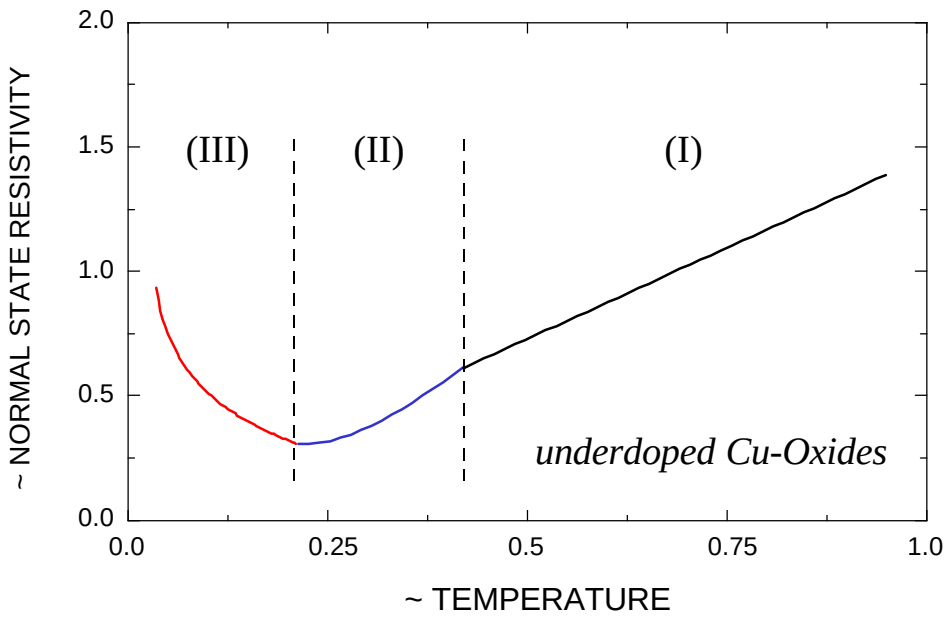

**Figure 2.** Sketch of the normal state resistivity of underdoped Cu-Oxide HTSC.

| | Region III | Region II | Region I |
|---|---|---|---|
| **Temperature dependence** | $r(T) = r_0 + r_1 \ln\left(\frac{1}{T}\right)$ (3) | $r(T) = r_0 + C\frac{T}{\Delta}\exp\left(-\frac{\Delta}{T}\right)$ (1) | $r_{2D}^{-1}(T) = s_{2D}(T)\big\vert_{L_f = x_m} \propto \frac{J}{T}$ |
| **Field dependence** | ? | $r(B) \propto \frac{T_{fix}}{\Delta_0(1-a\ B)} \exp\left(-\frac{\Delta_0(1-a\ B)}{T_{fix}}\right)$ (2) | *nearly field independent* |

**Table 1.** Overview of the proposed model which describes the normal state properties of underdoped Cu-Oxide superconductors. Details can be found elsewhere [5,6] Here $\Delta$ is the spin gap, $\Delta_0$ the zero field spin gap, $T_{fix}$ is a fixed temperature at which the field dependence is measured, $\alpha$ is a proportionality factor and $J$ is the exchange interaction energy.

In the following sections we will concentrate on the **regions II and III**. We will first discuss the spin gap effects by comparing the results of the underdoped HTSC with spin ladders, which are well studied spin gap systems. In a later stadium we will go to the low temperature range, where we observe the normal state resistivity divergence logarithmically. The latter could be described as a result of the interaction between disorder and stripes[7], which could interrupt the conduction paths by forcing the carriers to move from stripe to stripe.

## 2.1 *Comparison between the spin ladders and underdoped HTSC in Region II*

Figure 3 shows the temperature dependence of both the underdoped $La_{1.9}Sr_{0.1}CuO_4$ and the pressurized spin ladder compound $Sr_{2.5}Ca_{11.5}Cu_{24}O_{41}$[4]. At temperatures $T/T_0 < 0.6$, these curves do fit nicely each other. In the spin ladder compound, it is known that the transport is dominated by the one dimensional magnetic scattering. In this case, the temperature dependence of the resistivity ρ(T) can be described by Eq 1 (*see table 1*). If we apply Eq. 1, to fit the ***spin ladder data*** in figure 3, in the temperature range $0.2 < T/T_0 < 1$, we obtain the fitting parameters $r_0$, $C$ and $D$. In the model, the expected residual resistance $\rho_0$ is given by: $\rho_0$ = 2 Ñ $b^2$ ∆€/ ($e^2$ a π€$J_{//}$). Taking for the inter-stripe distance $b \sim 2a \sim 7.6$ Å, the spin gap $D \approx 200$ K and the exchange energy $J_{//} \sim 1400$ K (the normal value for the $CuO_2$ planes), the resistivity $r_0 \approx 0.5\cdot 10^{-4}$ Ωcm is in good agreement with $r_0 \approx 0.83\cdot 10^{-4}$ Ωcm found from the fit. The fitted gap $D \approx 216$ K (at 8 GPa) is in the vicinity of $D \approx 320$ K which is determined for the undoped spin ladder $SrCu_2O_3$ from inelastic neutron scattering experiments. In doped systems, however, it's natural to expect a reduction of the spin gap. Therefore the difference between the fitted value (216 K) and the one measured in an undoped system (320 K) seems to be quite fair.

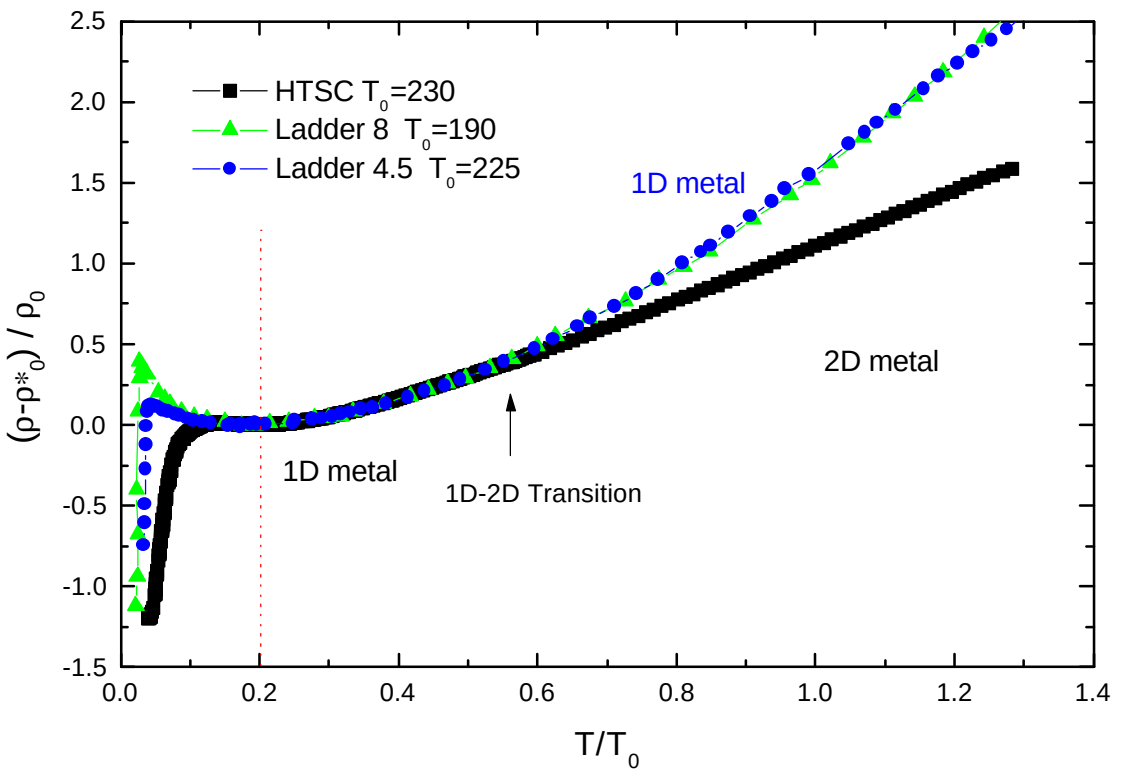

**Figure 3.** Relative resistivity $(\rho-\rho_0^*)/\rho_0$ versus temperature $T/T_0$. The HTSC and ladder compound scale with each other for $T/T_O < 0.6$.

If we also apply Eq. 1, to fit the ***underdoped HTSC data*** in figure 3, in the temperature range $0.2 < T/T_0 < 0.6$, we find similar fitting parameters $D \approx 183$ K and $r_0 \approx 3.3\cdot 10^{-4}$ Ωcm. The larger value of $\rho_0$ also seems to indicate a larger inter-stripe distance ( b ≈ 70 Å).

From the magnetoresistance in region II, we can also estimate the field dependence of the spin gap. In this region we have fitted Eq. 2 (*see table 1*) for the magnetoresistance. The motivation for the use of Eq. 2 is given by the fact that the field dependence of the spin gap can be expressed as a simple linear function of the field $\Delta(B) = \Delta_0$ (1 - α€B). The values of $\rho_0$, C and $\Delta_0$ are ***identical*** to the ones obtained using Eq. 1, from the fit of the temperature dependence. The temperature $T_{fix}$, is the temperature at which the magnetoresistance is measured, and thus, the ***only real fitting parameter*** is the slope α. The results of such a fit can be seen in figure 4a. The consequences for the field dependence of the spin gap are shown in figure 4b. The spin gap closes as the field increases, and the lower the temperature, the faster the gap closes.

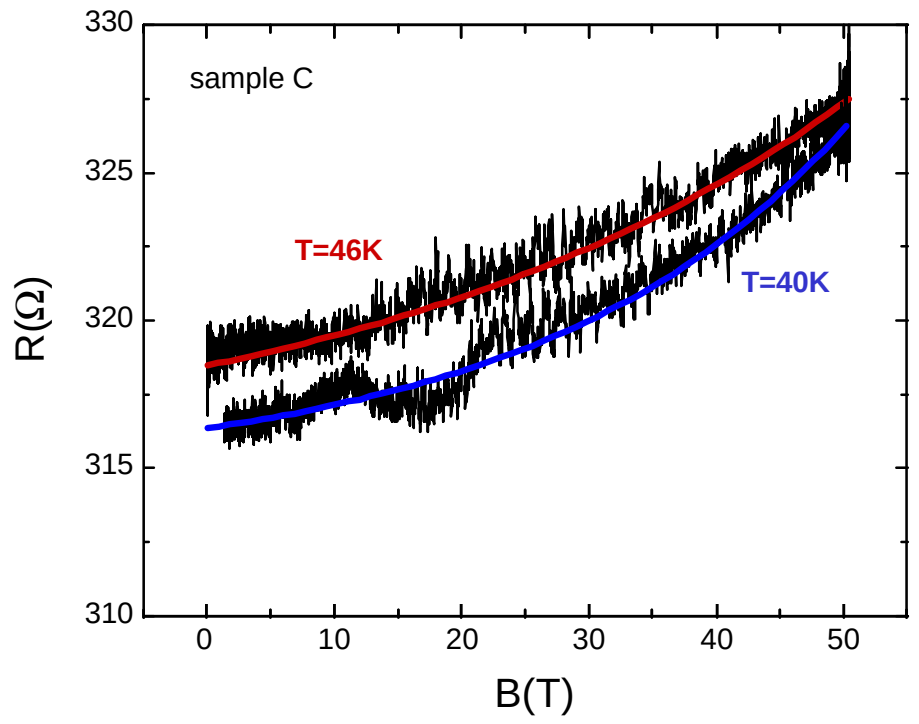

**Figure 4.a**. Magnetoresistance of sample C, measured at T=40K and T=46 K, up to B=50T. The lines are fits using Eq. 4.

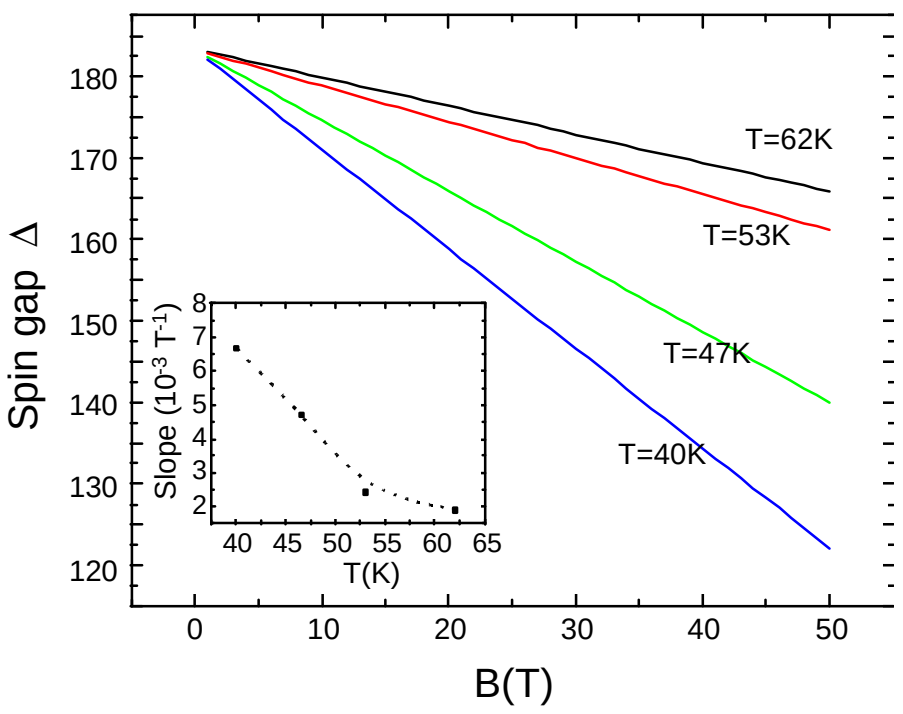

**Figure 4.b**. Field dependence of the spin gap, as obtained from fitting Eq. 4 to the experimental data of Fig 4.a

### 2.2 *The logarithmic divergence (Region III)*

Using pulsed magnetic fields, *r(H)* was measured up to 50 T. A typical set of data $\rho_{ab}$ obtained on sample A ($T_c$ ~ 45 K) is given in the inset of figure 5. From this type of experiments, we have taken the $m_oH$ = 50 T data to reveal the normal state *r(T)* below $T_c$ (figure 5). It should be mentioned that these $\rho_{ab}$ data also contain the normal state magneto-resistivity contribution which is not to be ignored completely, as can be seen from the data presented above (especially for the measurements above $T_c$ ($T$ = 72.7 K, $T$ = 61.6 K). Considering the temperature dependent *r(T),* it can be noted that the general trend of *r(T)* above $T/T_c \approx 0.2$ is extrapolated to the temperatures below $T/T_c \approx 0.2$. For the underdoped LSCO sample ($SrTiO_3$ substrate) above $T/T_c \approx 0.2$ one can already see a tendency towards an increasing resistivity at low temperatures. For the samples prepared on $SrLaAlO_4$, a 2D-metallic like behavior is always observed above $T/T_c$» 0.2 which seems to saturate below $T_c$.

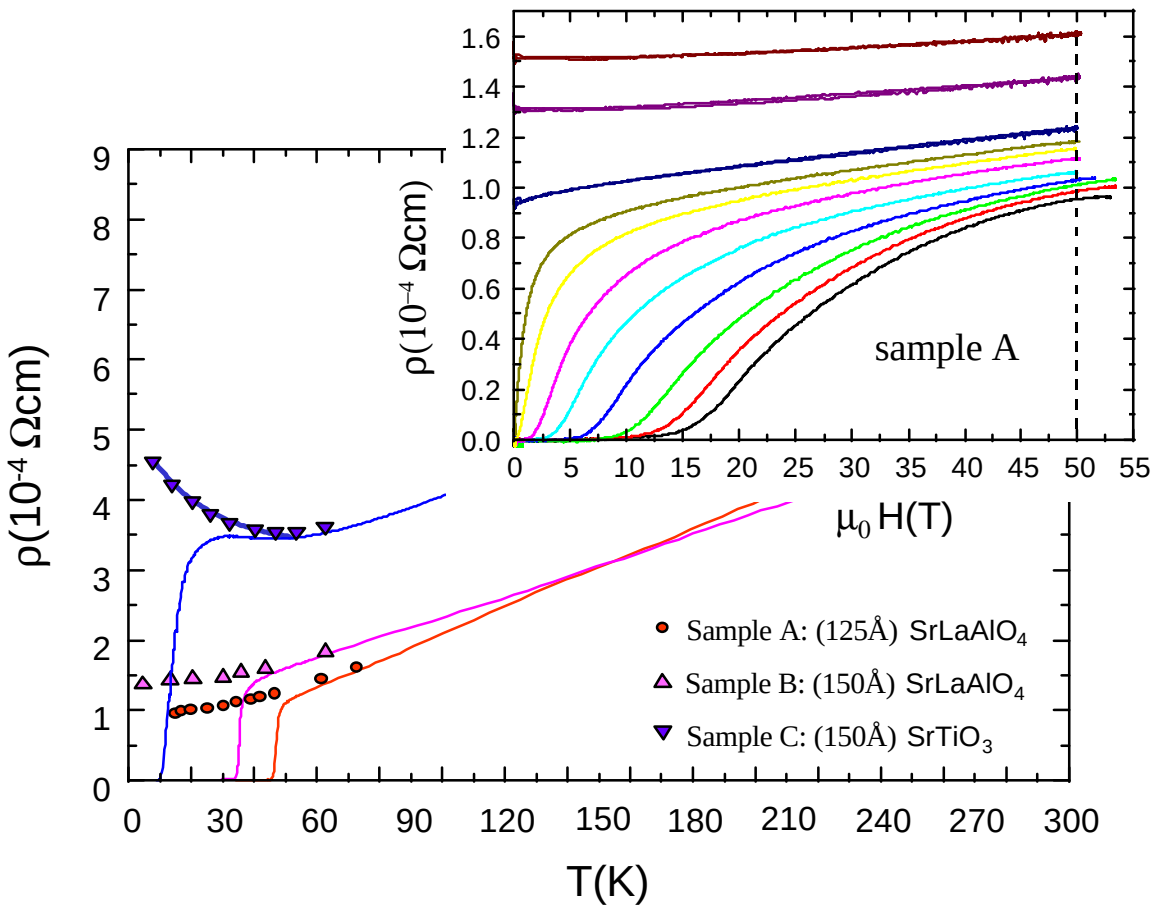

**Figure 5.** The temperature dependence of the resistivity $\rho_{ab}$ for the different $La_{1.9}Sr_{0.1}CuO_4$ samples (A,B and C). The data points indicated by the symbols are the $B$ = 50 T resistivity values. **Insert:** Magneto-resistivity of sample A at different temperatures (from top to bottom: $T$ = 72.7 K, $T$ = 61.6 K, $T$ = 46.5 K, $T$ = 41.9 K, $T$ = 39.0 K, $T$ = 34.4 K, $T$ = 30.2 K, $T$ = 25.1 K, $T$ = 19.9 K, $T$ = 16.9 K and $T$ = 14.7 K).

For the near optimum doped materials, we find a saturating resistance at lowering temperature, like the residual impurity resistance for normal metals. In the underdoped LSCO compound, however, we find a divergence of the resistance in lowering the temperature. This divergence can be best fitted by the logarithmic divergence, Eq. *3 (see table 1*). One might interpret such a logarithmic increase of the resistance $\rho_{ab}$ as a result of the destruction of the 1D-stripes by disorder. In such disordered stripe structures, defects prohibit the conduction along the stripes, and force the effective recovery of the 2D regime.

**Acknowledgments.** This work is supported by the Flemish FWO/GOA, the Flemish IWT, the Belgian IUAP Programs and the Flemish - Chinese bilateral agreement N° *BIL97/35* as well as the Swiss NSF project *2029-050538.97*.

## Corresponding author:

Johan Vanacken
Laboratorium voor Vaste-Stoffysica en Magnetisme
Katholieke Universiteit Leuven
Celestijnenlaan 200 D
B-3001 Heverlee
Belgium

tel: (32) 16 32 71 98
fax: (32) 16 32 79 83
e-mail: johan.vanacken@ fys.kuleuven.ac.be
http://www.fys.kuleuven.ac.be/vsm/vsm.html